\documentclass[12pt]{article}
\usepackage{cite}
\usepackage{enumerate}
\usepackage{ifpdf}
\usepackage{ae}
\usepackage[T1]{fontenc}
\usepackage[ansinew]{inputenc}
\usepackage{amsmath}
\usepackage{relsize}
\usepackage{amssymb}
\usepackage{tikz}
\usepackage{graphicx}
\usepackage{color}
\definecolor{darkblue}{cmyk}{0.9,0.9,0,0}
\definecolor{darkgreen}{rgb}{0,0.55,0}
\usepackage[colorlinks=true,linkcolor=darkblue,citecolor=darkblue,urlcolor=darkblue]{hyperref}
\usepackage{epsfig}
\usepackage{epstopdf}
\usepackage{graphicx}

\makeatletter
\long\def\@makecaption#1#2{
  \vskip\abovecaptionskip
  \sbox\@tempboxa{{\captionfonts #1: #2}}
  \ifdim \wd\@tempboxa >\hsize
    {\captionfonts #1: #2\par}
  \else
    \hbox to\hsize{\hfil\box\@tempboxa\hfil}
  \fi
  \vskip\belowcaptionskip}
\makeatother

\newcommand{\beq}{\begin{equation}}
\newcommand{\eeq}{\end{equation}}
\newcommand{\beqy} {\begin{eqnarray}}
\newcommand{\eeqy} {\end{eqnarray}}
\newcommand{\bsmat}{\begin{smallmatrix}}
\newcommand{\esmat}{\end{smallmatrix}}
\newcommand{\bmat}{\begin{matrix}}
\newcommand{\emat}{\end{matrix}}

\def\({\left(}
\def\){\right)}
\def\[{\left[}
\def\]{\right]}

\def\<{\langle}
\def\>{\rangle}

\usepackage{varioref}
\usepackage{makeidx}
\makeindex

\usepackage[english]{babel}
\usepackage{caption}
\usepackage{subfig}

\usepackage{tabularx}
\begin{document}

\thispagestyle{empty}

\renewcommand{\thefootnote}{\fnsymbol{footnote}}
\setcounter{page}{1}
\setcounter{footnote}{0}
\setcounter{figure}{0}

\begin{center}
{\LARGE\textbf{\mathversion{bold}
Generalized bootstrap equations \\ for ${\cal N}=4$  SCFT}}
\vspace{1.0cm}

\textrm{\Large Luis F. Alday and Agnese Bissi}
\\ \vspace{1.2cm}

\textit{Mathematical Institute, University of Oxford,}  \\
\textit{Radcliffe Observatory Quarter, Oxford, OX2 6GG, UK} \\
\vspace{5mm}

\par\vspace{1.5cm}

\textbf{Abstract}\vspace{2mm}
\end{center}

\noindent
We study the consistency of four-point functions of half-BPS chiral primary operators of weight $p$ in four-dimensional ${\cal N}=4$ superconformal field theories. The resulting conformal bootstrap equations impose non-trivial bounds for the scaling dimension of unprotected local operators transforming in various representations of the R-symmetry group. These bounds generalize recent bounds for operators in the singlet representation, arising from consistency of the four-point function of the stress-energy tensor multiplet.

\vspace*{\fill}

\setcounter{page}{1}
\renewcommand{\thefootnote}{\arabic{footnote}}
\setcounter{footnote}{0}

\newpage

 \def\nref#1{{(\ref{#1})}}

\section{Introduction}

In recent years there has been substantial progress in our understanding of conformal field theories (CFT) in dimensions higher than two. In general such theories do not admit a Lagrangian description, so one has to resort to consistency conditions arising from conformal symmetry, unitarity, crossing symmetries and the properties of the operator product expansion (OPE). This is the idea of the conformal bootstrap program. In the simplest set-up one considers the four-point correlator of a scalar field $\phi$ of dimension $d$. Conformal symmetry implies
\begin{equation}
\langle \phi(x_1)\phi(x_2)\phi(x_3)\phi(x_4) \rangle = \frac{g(u,v)}{x_{12}^{2d}x_{34}^{2d}}
\end{equation}
where we have introduced the cross-ratios $u=(x_{12}^2 x_{34}^2)/(x_{13}^2x_{24}^2)$ and $v=(x_{14}^2 x_{23}^2)/(x_{13}^2x_{24}^2)$. By considering the OPE $\phi(x_1) \times \phi(x_2)$ we can decompose the four-point function into conformal blocks
\begin{equation}
\label{opedec}
g(u,v) = 1 +\sum_{\Delta,\ell} a_{\Delta,\ell} g_{\Delta,\ell}(u,v)
\end{equation}
where we have singled out the contribution from the identity operator. The sum runs over the tower of conformal primaries present in the OPE ( ${\cal O}_{\Delta,\ell} \in \phi \times \phi$ ) and $\Delta$ and $\ell$ denote the dimension and the spin of the intermediate primary. $a_{\Delta,\ell}$ denotes the square of the structure constants and is non-negative due to unitarity. The conformal blocks $g_{\Delta,\ell}(u,v)$ repack the contribution of all descendants of a given primary and are fixed by conformal symmetry. They depend only on the spin and dimension of the primary. Crossing-symmetry of the four-point function 
\begin{equation}
 \frac{g(u,v)}{x_{12}^{2d} x_{34}^{2d}}= \frac{g(v,u)}{x_{23}^{2d} x_{14}^{2d}}~ \rightarrow ~v^d g(u,v) = u^d g(v,u)
\end{equation}
together with associativity of the OPE imply the conformal bootstrap equation
\begin{eqnarray}
&\sum_{\ell,\Delta} a_{\Delta,\ell} F_{\Delta,\ell}(u,v) =1,~~~~~~a_{\Delta,\ell}  \geq 0\\
&F_{\Delta,\ell}(u,v) \equiv  \frac{v^d g_{\Delta,\ell}(u,v)- u^d g_{\Delta,\ell}(v,u)}{u^d-v^d} \nonumber
\end{eqnarray}
As shown in \cite{Rattazzi:2008pe}, the conformal bootstrap equation can be used to put upper bounds to the dimensions of leading twist primary operators appearing in the OPE $\phi \times \phi$. 

One can also analyze CFT's with a continuous global symmetry group \cite{Rattazzi:2010yc}. In this case the natural starting point is the four-point correlator
\begin{equation}
\langle \phi \phi \phi^\dagger \phi^\dagger \rangle
\end{equation}
where the scalar primary operator $\phi$ transforms in a given representation ${\cal R}$ of the global symmetry group. For instance, for $SO(N)$ global symmetry and $\phi$ transforming in the fundamental representation, the OPE $\phi \times \phi$ contains states transforming as singlets $S$, symmetric traceless tensors $T_{(ij)}$ or antisymmetric tensors $A_{[ij]}$. Consequently the conformal bootstrap equation has a vector structure mixing these three components:

\begin{equation}
\label{global}
\sum_{\Delta, \ell} a_{\Delta, \ell}^{S} \left(\begin{matrix}0\\ F_{\Delta, \ell}\\ H_{\Delta, \ell} \end{matrix}\right)+\sum_{\Delta, \ell} a_{\Delta, \ell}^{T} \left(\begin{matrix} F_{\Delta, \ell}\\ \left( 1-\frac{2}{N} \right)F_{\Delta, \ell}\\ -\left( 1+\frac{2}{N}\right)H_{\Delta, \ell}\end{matrix}\right)+\sum_{\Delta, \ell} a_{\Delta, \ell}^{A} \left(\begin{matrix} - F_{\Delta, \ell}\\ F_{\Delta, \ell} \\ -H_{\Delta, \ell}\end{matrix}\right)=\left(\begin{matrix}0\\ 1\\ -1\end{matrix}\right)
\end{equation}
where we have introduced $H_{\Delta,\ell}(u,v) \equiv  \frac{v^d g_{\Delta,\ell}(u,v)+ u^d g_{\Delta,\ell}(v,u)}{u^d+v^d}$. Again, the conformal bootstrap equations can be used to put bounds on the dimensions of operators appearing in the OPE, see {\it e.g.} \cite{Rattazzi:2010yc,Poland:2011ey,Kos:2013tga}.

Supersymmetric conformal field theories (SCFT) play a predominant role in theoretical physics. Very recently the conformal bootstrap program has been extended to four-dimensional ${\cal N}=4$ SCFT \cite{Beem:2013qxa} \footnote{See \cite{Poland:2010wg,Vichi:2011ux,Poland:2011ey,Berkooz:2014yda,Khandker:2014mpa} for extensions to four-dimensional ${\cal N}=1$ SCFT. }. In this case the energy momentum tensor lies in a half-BPS multiplet whose superconformal primary is a scalar operator of dimension two, which transforms in the $[0,2,0]$ representation of the $SU(4)$ R-symmetry group. The natural object to consider is the four-point function of such scalar operator
\begin{equation}
\langle {\cal O}^{[2]}(x_1){\cal O}^{[2]}(x_2){\cal O}^{[2]}(x_3){\cal O}^{[2]}(x_4) \rangle = \frac{{\cal G}(u,v)}{x_{12}^4x_{34}^4}
\end{equation}
This correlator decomposes into six channels, corresponding to the possible representations of the intermediate states
\begin{equation}
[0,2,0] \times [0,2,0] = [0,0,0]+[1,0,1]+[0,2,0]+[2,0,2]+[1,2,1]+[0,4,0]
\end{equation}
The contribution from each channel could be written as a sum over {\it conformal} primaries with the corresponding conformal blocks, as in (\ref{opedec}). On the other hand each intermediate operator belongs to a particular superconformal multiplet. Hence, the correlator can also be written as a sum over {\it superconformal} primaries. For the present case only superconformal primaries transforming in the singlet representation $[0,0,0]$ belong to long unprotected multiplets \cite{Arutyunov:2001qw,Eden:2001ec}, while the contribution from other superconformal primaries is fixed by the superconformal Ward identities \cite{Dolan:2001tt}. The conformal bootstrap equation takes the final form \cite{Beem:2013qxa} 

\begin{equation}
\label{super}
\sum_{\substack{\ell=0,2,..., \\ \Delta \geq \ell+2}} a_{\Delta,\ell} F_{\Delta, \ell}(u,v)=F^{short}(u,v,c)
\end{equation}
where the sum runs over superconformal primaries singlet of $SU(4)$. $F^{short}(u,v,c)$ arises from short and semi-short contributions to the correlator and depends only on the central charge $c$ of the theory, which appears in the OPE of two stress tensors. Since operators in short representations may combine into long representations at the unitarity bound, there is an ambiguity when computing $F^{short}(u,v,c)$. There is a canonical choice for which all the coefficients $a_{\Delta,\ell}$ are non-negative. As shown in \cite{Beem:2013qxa} this equation can be used to find upper bounds for the scaling dimensions of leading twist operators transforming in the singlet representation of $SU(4)$, such as the Konishi operator. 

The aim of this paper is to study the consistency of more general four-point functions in four-dimensional ${\cal N}=4$ SCFT. More precisely we will study four-point correlation functions of identical chiral primary half-BPS operators transforming in the $[0,p,0]$ representation of the $SU(4)$ R-symmetry group: 
\begin{equation}
\langle {\cal O}^{[p]}(x_1)  {\cal O}^{[p]}(x_2)  {\cal O}^{[p]}(x_3) {\cal O}^{[p]}(x_4)\rangle
\end{equation}
The constraints of superconformal invariance on these correlators were studied in detail in \cite{Heslop:2002hp,Arutyunov:2003ad,Dolan:2004mu,Nirschl:2004pa,Dolan:2004iy}. The correlation function can be decomposed into $\frac{(p+1)(p+2)}{2}$ channels but again, only a restricted subset ($p(p-1)/2$ of them) contains unprotected superconformal primary operators. In the next section we derive the conformal bootstrap equations arising from crossing-symmetry of such correlation functions. They are given by  $p(p-1)/2$ coupled equations and have the form of bootstrap equations for CFT's with global symmetry, see eq. \eqref{oureqs}. In addition to the central charge, the right hand side of these equations depends on additional information about the SCFT, namely extra parameters that arise in the OPE of symmetric-traceless tensors of rank $p$. In section three we use these equations to find rigorous bounds for the anomalous dimensions of superconformal primaries of ${\cal N}=4$ SYM with gauge group $SU(N)$ as a function of the rank of the gauge group. We focus in the case $p=3$ and find bounds for operators transforming in the representations $[1,0,1]$ and $[0,2,0]$ of the R-symmetry group. We end up with some conclusions. Finally, several technical details are discussed in the appendices.

\section{Generalized bootstrap equations}
\label{booteq}

The superconformal algebra of four-dimensional ${\cal N}=4$ SCFT is $PSU(2,2|4)$. This algebra contains a $SU(4)$ R-symmetry group. The energy-momentum tensor lies in a half-BPS multiplet whose superconformal primary is a scalar operator transforming in the $[0,2,0]$ representation of the R-symmetry group. This scalar operator is part of a family of half-BPS scalar operators ${\cal O}^{[p]}$ of dimension $p$, transforming in the $[0,p,0]$ representation \footnote{More generaly we consider $\varphi^{[p]}=\varphi_{r_1...r_p} t_{r_1}...t_{r_p}$ where $\varphi_{r_1...r_p}$ is a symmetric traceless tensor field.}

\begin{equation}
{\cal O}^{[p]}(x,t)= t_{r_1}\dots t_{r_p} ~\textrm{Tr} \left(\Phi^{r_1} \cdots \Phi^{r_p} \right)
\end{equation}
with  $t$ a complex six-dimensional null vector ( $t \cdot t =0$ ) and $r_i=1,...,6$. The correlator of four identical such operators can be written as \cite{Dolan:2004iy}
\begin{equation}
\langle {\cal O}^{[p]}(x_1,t_1)  {\cal O}^{[p]}(x_2,t_2)  {\cal O}^{[p]}(x_3,t_3) {\cal O}^{[p]}(x_4,t_4)\rangle= \left(\frac{t_1\cdot t_2 \,t_3 \cdot t_4}{x_{12}^2 x_{34}^2} \right) ^p\mathcal{G}^{(p)}(u,v,\sigma, \tau)
\end{equation}
where $u$ and $v$ are conformal invariant cross-ratios while $\sigma$ and $\tau$ are $SU(4)$ invariants
\begin{equation}
\begin{gathered} 
u=\frac{x_{12}^2 x_{34}^2}{x_{13}^2 x_{24}^2}  \qquad v=\frac{x_{14}^2 x_{23}^2}{x_{13}^2 x_{24}^2}  \\ \sigma=\frac{t_1\cdot t_3\, t_2\cdot t_4}{t_1\cdot t_2\, t_3 \cdot t_4} \qquad \tau=\frac{t_1\cdot t_4\, t_2\cdot t_3}{t_1\cdot t_2\, t_3 \cdot t_4}
\end{gathered} 
\end{equation}
$\mathcal{G}^{(p)}(u,v,\sigma, \tau)$  is a polynomial in $\sigma$ and $\tau$ of degree $p$ and can be decomposed into $\frac{(p+1)(p+2)}{2}$ contributions corresponding to the different $SU(4)$ representations in the tensor product
\begin{equation}
[0,p,0] \times [0,p,0]=\sum_{k=0}^p\sum_{q=0}^{p-k}[q,2p-2q-2k,q]
\end{equation}
Each of these contributions can be expanded in conformal partial waves, corresponding to conformal primary operators with dimensions $\Delta$ and spin $\ell$ transforming in the appropriate representation. Superconformal symmetry implies that each conformal primary belongs to a given supermultiplet, with a corresponding superconformal primary (which does not necessarily transform in the same $SU(4)$ representation). In general it is quite involved to separate the contributions  in the conformal partial wave expansion of descendant operators from superconformal primary operators. As explained in detail in \cite{Nirschl:2004pa,Dolan:2004iy} this can be done by solving explicitly the superconformal Ward identities. More precisely, superconformal Ward identities dictate the decomposition of $\mathcal{G}(u,v,\sigma, \tau)$  in terms of long multiplets, containing all the dynamical non-trivial information, and short and semi-short multiplets, which are fully determined by symmetries and the free field theory results. Hence $\mathcal{G}$ can be expressed as follows
\begin{equation}
\label{h}
\mathcal{G}\left(z, \bar{z}, \alpha, \bar{\alpha} \right)=k+ \mathcal{G}_{\hat{f}}+(\alpha  z-1) (\bar{\alpha} z-1) (\alpha \bar{z}-1) (\bar{\alpha}
  \bar{z}-1)\mathcal{H}\left(z, \bar{z}, \alpha, \bar{\alpha} \right)
\end{equation}
where we have suppressed the index $p$ and have introduced the variables
\begin{equation}
\label{xxb}
\begin{gathered}
u=z \bar{z} \qquad v=(1-z)(1-\bar{z})\\
\sigma=\alpha \bar{\alpha} \qquad \tau=(1-\alpha)(1-\bar{\alpha})
\end{gathered}
\end{equation}
The function $\mathcal{G}_{\hat{f}}$ depends only on free theory results while $\mathcal{H}$ includes dynamical effects,
\begin{equation}
\mathcal{G}_{\hat{f}}=\frac{(\bar{\alpha} z-1) (\alpha  \bar{z}-1) (F(z,\alpha
   )+F(\bar{z},\bar{\alpha}))-(\alpha  z-1) (\bar{\alpha}
   \bar{z}-1) (F(z,\bar{\alpha})+F(\bar{z},\alpha ))}{(\alpha
   -\bar{\alpha}) (z-\bar{z})}-2 k
\end{equation}
and
\begin{equation}
F(z,\alpha)=\left(\alpha -\frac{1}{z}\right) \hat{f}(z,\alpha )+k \qquad \hat{f}(z,\alpha)=\mathcal{G}_0\left(z, \bar{z}, \alpha, \bar{\alpha} \right)_{\bar{\alpha}=\frac{1}{\bar{z}}}\qquad k=\hat{f}(z,z)
\end{equation}
where $\mathcal{G}_0$ denotes the tree level four-point function. As already mentioned $\mathcal{H}\left(z, \bar{z}, \alpha, \bar{\alpha} \right)$ encodes the non-trivial, unprotected, information of the four-point function. It turns out it receives contributions only from a restricted set of representations ($p(p-1)/2$ of them) and can be written as
\begin{eqnarray}
\mathcal{H}(z,\bar{z},\alpha, \bar{\alpha})&=&\sum\limits_{\substack{0 \leq m \leq n \leq p-2}}\mathcal{H}^{[nm]}(z,\bar{z}) Y_{nm}(\alpha, \bar{\alpha})
\label{pwH}\\
\label{Hoverl}
\mathcal{H}^{[nm]}(z,\bar{z})&=&\sum\limits_{\substack{\Delta, \ell}} A^{[nm]}_{\Delta,\ell}\, (z \bar{z})^{\frac{1}{2}\left(\Delta-\ell \right)} G_{\Delta+4}^{(\ell)}(z,\bar{z})
\end{eqnarray}
where we have introduced a short-hand notation $[nm] \equiv [n-m,2m,n-m]$ for $SU(4)$ representations. The harmonic polynomials $Y_{n m}(\alpha, \bar{\alpha})$ encode the dependence on the $SU(4)$ invariants and have an explicit definition in terms of Legendre polynomials \cite{Dolan:2004iy}:
\begin{equation}
Y_{nm}(\alpha, \bar{\alpha}) = \frac{P_{n+1}(2\alpha-1) P_{m}(2\bar \alpha -1)-P_{m}(2\alpha-1) P_{n+1}(2\bar \alpha -1)}{2(\alpha-\bar \alpha)}
\end{equation}
The sum over the spin in (\ref{Hoverl}) runs over even/odd spins if $n+m$ is even/odd. $G_{\Delta}^{(\ell)}(z,\bar{z})$ denote the four-dimensional conformal blocks, given by
\begin{equation}
G_{\Delta}^{(\ell)}(z,\bar{z})=\frac{1}{z-\bar{z}}\left( \left( -\frac{1}{2}z\right)^\ell z \kappa_{\Delta+\ell}(z)\kappa_{\Delta-\ell-2}(\bar{z})- (z \leftrightarrow \bar{z}) \right)
\end{equation}
with $\kappa_{\beta}(z)=\ _2F_1(\frac{\beta}{2},\frac{\beta}{2},\beta, z)$.  Unitarity requires that only contributions for $\Delta \geq 2n+ \ell+2$ arise and that the coefficients $A^{[n m]}_{\Delta,\ell}$ are non-negative \footnote{Since we would like to interpret them as the square of the structure constants of two half-BPS operator transforming in the $[0,p,0]$ and one superconformal primary operator of dimension $\Delta$ and spin $\ell$ transforming in the $[n-m,2m,n-m]$.}. This is not automatic. On the other hand, there is an ambiguity since a long multiplet decomposes into semi-short multiplets at the unitary threshold. This ambiguity allows to letting $A^{[nm]}_{\Delta, \ell} \to a^{[nm]}_{\Delta, \ell}$ where now
\begin{eqnarray}
\mathcal{H}(z,\bar{z},\alpha, \bar{\alpha})&=&\sum\limits_{\substack{0 \leq m \leq n \leq p-2}}\hat{\mathcal{H}}^{[nm]}(z,\bar{z}) Y_{nm}(\alpha, \bar{\alpha})
\label{pwhH}\\
\hat{\mathcal{H}}^{[nm]}(z,\bar{z})&=&\sum\limits_{\substack{\Delta, \ell}} a^{[nm]}_{\Delta,\ell}\, (z \bar{z})^{\frac{1}{2}\left(\Delta-\ell \right)} G_{\Delta+4}^{(\ell)}(z,\bar{z})+F_{(p)}^{[nm]}(z,\bar{z}) \label{Hhatdec}
\end{eqnarray}
The functions $F_{(p)}^{[nm]}(z,\bar{z})$ contain only contributions from short and semi-short multiplets for each specific $SU(4)$ representation and do not depend on the coupling constant. There is a canonical choice which makes the coefficients $a^{[nm]}_{\Delta,\ell}$ non-negative and the expansion consistent with unitarity. This choice was explicitly worked out in \cite{Dolan:2004iy} for $p=2,3,4$ and is reproduced in appendix \ref{fshorts} for $p=3$. The coefficients $a^{[nm]}_{\Delta,\ell}$ are then interpreted as the square of the structure constants of two half-BPS operator transforming in the $[0,p,0]$ and the superconformal primary operator of dimension $\Delta$ and spin $\ell$ transforming in the $[n-m,2m,n-m]$.

Crossing symmetry requires invariance of the four-point function under exchanging $(x_1,t_1)$ with $(x_3,t_3)$. This entails $u\to v$, $v \to u$, $\sigma \to \frac{\sigma}{\tau}$ and $\tau \to \frac{1}{\tau}$ at the level of cross ratios and implies
\begin{equation}
\label{cross}
\mathcal{G}\left(z, \bar{z}, \alpha, \bar{\alpha} \right)= (1-\alpha)^p(1-\bar{\alpha})^p \left(\frac{z \bar{z}}{(1-z)(1-\bar{z})} \right)^p \mathcal{G}\left(1-z, 1-\bar{z}, \frac{\alpha}{1-\alpha}, \frac{\bar{\alpha}}{1-\bar{\alpha}}\right)
\end{equation}
When substituting \eqref{h} into \eqref{cross} one obtains an equation for the function $\mathcal{H}\left(z, \bar{z}, \alpha, \bar{\alpha} \right)$. Plugging in this equation the conformal partial wave decomposition \eqref{pwhH} and projecting over $SU(4)$ representations it is possible to write $\frac{p(p-1)}{2}$ equations for combinations of $\hat{\mathcal{H}}^{[nm]}(z,\bar{z})$.

\bigskip

For $p=2$ only the singlet representation contributes to the conformal partial wave decomposition of $\hat{\mathcal{H}}^{[nm]}(z,\bar{z})$ and \eqref{cross} implies

\begin{equation}
\label{p2crossing}
u^2 \hat{\mathcal{H}}^{[00]}(v,u) -v^2 \hat{\mathcal{H}}^{[00]}(u,v)-(u-v)(a_2+a_1(u+v))=0 
\end{equation}
Using the decomposition \eqref{Hhatdec} this implies

\begin{equation}
\label{p2eq}
\sum\limits_{\substack{\Delta \geq \ell+2\\ \ell=0,2,\dots}} a_{\Delta, \ell}^{[00]} F_{\Delta, \ell}^{(2)}(u,v) =F_{short}(u,v)
\end{equation}
where $F^{short}_{(2)}(u,v)$ can be worked out explicitly from the formulae in  \cite{Dolan:2004iy} and we have introduced 
\begin{equation}
F_{\Delta, \ell}^{(p)}(u,v)=v^p u^{\frac{1}{2}\left(\Delta-\ell \right)} G_{\Delta+4}^{(\ell)}(u,v)-u^p v^{\frac{1}{2}\left(\Delta-\ell \right)} G_{\Delta+4}^{(\ell)}(v,u)
\end{equation}
Equations \eqref{p2crossing} and \eqref{p2eq} exactly agree with the equations found by \cite{Beem:2013qxa}. As already mentioned $F^{short}_{(2)}(u,v)$ does not depend on the coupling constant. It only depends on two factors $a_1$ and $a_2$ related to different topologies of free field theory graphs. $a_1$ corresponds to the disconnected diagram and we can choose a normalization such that $a_1=1$. With this normalization $a_2$ is the inverse of the central charge of the theory. Hence, the central charge is the only information about the SCFT that enters the bootstrap equation for the case $p=2$. 
\bigskip

For $p=3$ the representations that contribute to the conformal partial wave decomposition of $\hat{\mathcal{H}}^{[nm]}(z,\bar{z})$ are  $[0,0,0]$, $[1,0,1]$ and $[0,2,0]$. The crossing equation \eqref{cross} implies 

\begin{multline}
u^3 (\hat{\mathcal{H}}^{[00]}(v,u)-15\hat{\mathcal{H}}^{[10]}(v,u)+20\hat{\mathcal{H}}^{[11]}(v,u))-6 v^3 \hat{\mathcal{H}}^{[00]}(u,v)-u^2 (a_1 u (4 u-3)+a_3)+3 a_1 u^3
   v  \\  +a_1(7 u+12) v^3-6 a_1 v^4+a_2 \left(-6 u^3-u^2 (v+2)+u (v-4) (v-1)+6 v (v+1)^2\right)+6 a_3 v^2=0, \notag
\end{multline}
\begin{multline}
u^3 (-\hat{\mathcal{H}}^{[00]}(v,u)+3 \hat{\mathcal{H}}^{[10]}(v,u)+4\hat{\mathcal{H}}^{[11]}(v,u))-6 v^3\hat{\mathcal{H}}^{[10]}(u,v)+u^2 (a_1 u+2 a_2 (u+1)+a_3) \\
 -v
   \left(a_1 u^3+a_2 \left(u^2+u-2\right)\right)+v^3 (a_1 (3 u+2)-2 a_2)-2 a_1 v^4+a_2 u v^2=0, \notag
\end{multline}   
\begin{multline}   
u^3 (\hat{\mathcal{H}}^{[00]}(v,u)+3\hat{\mathcal{H}}^{[10]}(v,u)+2\hat{\mathcal{H}}^{[11]}(v,u))-6 v^3\hat{\mathcal{H}}^{[11]}(u,v)+u \left(a_1 \left(2 u^3-3 u^2
   (v+1)+v^3\right) \right.\\
\left.  -a_2 (v+2) (u-v+1)-a_3 u\right)=0, \notag
\end{multline}
where the factors $a_1,a_2,a_3$ correspond to different topologies of the graphs contributing to the tree-level answer. For ${\cal N}=4$ SYM with gauge group $SU(N)$ they are functions of the rank of the gauge group. For the conformal bootstrap analysis of the next section, it is important to compute them for finite rank. This is done in appendix \ref{colorfactors}. In order to find the conformal bootstrap equations we simply plug in the decomposition  \eqref{Hhatdec}.  $F_{(3)}^{[nm]}(z,\bar{z})$ receives specific contributions from short and semi-short multiplets and the explicit sums are given and performed in appendix \ref{fshorts}. The final equations can be written in a very elegant vector form
\begin{equation}
\label{oureqs}
\sum\limits_{\substack{\Delta \geq \ell+2\\ \ell=0,2,\dots}} a_{\Delta, \ell}^{[00]} \left(\begin{matrix}F_{\Delta, \ell}^{(3)}\\ 0\\ H_{\Delta, \ell}^{(3)}\end{matrix}\right)+\sum\limits_{\substack{\Delta \geq \ell+4\\ \ell=1,3,\dots}} a_{\Delta, \ell}^{[10]} \left(\begin{matrix}0\\ F_{\Delta, \ell}^{(3)}\\ 3H_{\Delta, \ell}^{(3)}\end{matrix}\right)+\sum\limits_{\substack{\Delta \geq \ell+4\\ \ell=0,2,\dots}} a_{\Delta, \ell}^{[11]} \left(\begin{matrix} 5 F_{\Delta, \ell}^{(3)}\\ F_{\Delta, \ell}^{(3)}\\ -4H_{\Delta, \ell}^{(3)}\end{matrix}\right)=\left(\begin{matrix} F^{1}_{short}(u,v)\\ F^{2}_{short}(u,v)\\ F^{3}_{short}(u,v)\end{matrix}\right)
\end{equation}
where we have introduced a new structure
\begin{eqnarray}
H_{\Delta, \ell}^{(p)}(u,v)&=&v^p u^{\frac{1}{2}\left(\Delta-\ell \right)} G_{\Delta+4}^{(\ell)}(u,v)+u^p v^{\frac{1}{2}\left(\Delta-\ell \right)} G_{\Delta+4}^{(\ell)}(v,u)
\end{eqnarray}
and $F^{1}_{short}(u,v)$, $F^{2}_{short}(u,v)$ and $F^{3}_{short}(u,v)$ are simple combinations of $F^{[00]}_3(u,v)$, $F^{[10]}_3(u,v)$ and $F^{[11]}_3(u,v)$. These equations have the same structure as the conformal bootstrap equations in the presence of global symmetries and explore the non-trivial R-symmetry structure of the theory. 

\bigskip

For $p=4$ and higher the structure is very much the same. In general we obtain  $\frac{p(p-1)}{2}$ coupled equations that can be written in a vectorial form. These equations will involve $F_{\Delta, \ell}^{(p)}(u,v)$ and $H_{\Delta, \ell}^{(p)}(u,v)$ on the left hand side, and complicated (but independent of the coupling) contributions on the right hand side. The left hand side can be readily computed as above. In order to compute the right hand side one needs the specific substractions to be done in order to render the decomposition consistent with unitarity. To the best of our knowledge this has been worked out only for $p=2,3,4$. Furthermore, the equations will depend on factors $a_1$ (which can always be set to one) and $a_2,a_3,$ etc. $a_2$ depends only on the central charge of the theory (see appendix \ref{colorfactors}). More precisely 

\begin{equation}
a_2 = \frac{p^2}{4 \, c} 
\end{equation}
where $c$ is the central charge, given by $c=\textrm{dim}~ G/4$ for ${\cal N}=4$ SYM with gauge group $G$. On the other hand $a_3,a_4,$ etc, carry extra information about the SCFT and distinguish between different SCFT's with the same central charge. 

\section{Numerical bounds}

\subsection{Setup}

In this section we study the consequences of the conformal bootstrap equations found above for the dimension of operators in ${\cal N}=4$ SYM with gauge group $SU(N)$. We will focus in the case $p=3$ which was the one worked out in detail but the extension to $p=4$ should be straightforward. The bootstrap equations \eqref{oureqs} have the following three-dimensional vector structure

\begin{equation}
\sum_{\Delta,\ell} a_{\Delta,\ell}^{[00]} \vec{V}^{[00]}_{\Delta,\ell}+\sum_{\Delta,\ell} a_{\Delta,\ell}^{[10]} \vec{V}^{[10]}_{\Delta,\ell}+\sum_{\Delta,\ell} a_{\Delta,\ell}^{[11]} \vec{V}^{[11]}_{\Delta,\ell}= \vec{F}_{short}
\end{equation}
With non-negative coefficients $a_{\Delta,\ell}^{{\cal R}}$. Unitarity demands the following lower bounds for the dimensions of the operators

\begin{eqnarray}
\Delta \geq \ell+2 ~~\textrm{for}~~[00],~~~~~\Delta \geq \ell+4 ~~\textrm{for}~~[10]~\textrm{and}~[11]
\end{eqnarray}
Bootstrap equations with this structure appear when studying CFT's with global symmetries and were analyzed in \cite{Rattazzi:2010yc}. A given spectrum can be ruled out if we can find a linear functional $\Phi: \vec{V} \rightarrow R~$ such that
\begin{equation}
\begin{aligned}
&\Phi ~\vec{V}^{[00]}_{\Delta,\ell} \geq 0,~~~~~\textrm{for}~a_{\Delta,\ell}^{[00]} \neq 0,~\ell=0,2,...\\
&\Phi ~\vec{V}^{[10]}_{\Delta,\ell} \geq 0,~~~~~\textrm{for}~a_{\Delta,\ell}^{[10]} \neq 0,~\ell=1,3,...\\
&\Phi ~\vec{V}^{[11]}_{\Delta,\ell} \geq 0,~~~~~\textrm{for}~a_{\Delta,\ell}^{[11]} \neq 0,~\ell=0,2,...\\
&\Phi ~\vec{F}_{short} < 0.
\end{aligned}
\end{equation}
In order to write down the explicit linear operator we introduce the following variables:

\begin{equation}
z= 1/2+a+b,~~~\bar{z}=1/2+a-b.
\end{equation}
The linear operator takes the form

\begin{equation}
\Phi^{(\Lambda)} \left(\begin{matrix} f_1(a,b)\\ f_2(a,b) \\ f_3(a,b) \end{matrix}\right) = \sum_{i,j=0}^{i+j=\Lambda} \left( \frac{\xi^{(1)}_{ij}}{i! j!} \partial_a^i \partial_b^j f_1(0,0)+ \frac{\xi^{(2)}_{ij}}{i! j!} \partial_a^i \partial_b^j f_2(0,0)+ \frac{\xi^{(3)}_{ij}}{i! j!} \partial_a^i \partial_b^j f_3(0,0) \right)
\end{equation}
In order to have a finite problem we have set a cut-off in the spin and the twist $\Delta-\ell$. This is then supplemented by asymptotic expressions, valid for large dimension. Furthermore, we have discretized the possible dimensions for each spin, with step $\delta \Delta = 1/25$. We have used a linear operator with a maximum of eleven derivatives, $\Lambda=11$. This gives a linear operator that depends on 63 parameters. The inequalities were generated with \texttt{Mathematica} and then analyzed with the \texttt{IBM ILOG CPLEX} optimizer and \texttt{Matlab} .

\subsection{Results}

The conformal bootstrap equations found in this paper can be used to put upper bounds to the dimensions of leading twist superconformal primary operators in long multiplets, transforming in the representations $[0,0,0],[1,0,1]$ and $[0,2,0]$ of the R-symmetry group. These bounds are non-perturbative and depend on the rank of the gauge group through the color factors
\begin{equation}
a_1 = 9 (N^2-1)^2(N-\frac{4}{N})^2,~~~a_2=\frac{9}{N^2-1} a_1,~~~a_3=162(N^2-1)\frac{48-16 N^2+N^4}{N^2}
\end{equation}
where we are free to rescale them by an overall factor. From these expressions one can see that the conformal partial wave expansion is consistent with unitarity for $N \geq 3$, hence we will restrict to this range. For leading twist operators in the singlet representation, of the schematic form $\textrm{Tr}~\Phi^I D^\ell \Phi^I$,~ $\ell=0,2,...$, we have found bounds consistent with \cite{Beem:2013qxa} but much less constraining. 

\begin{figure}[h!]
\centering
\includegraphics[width=4.5in]{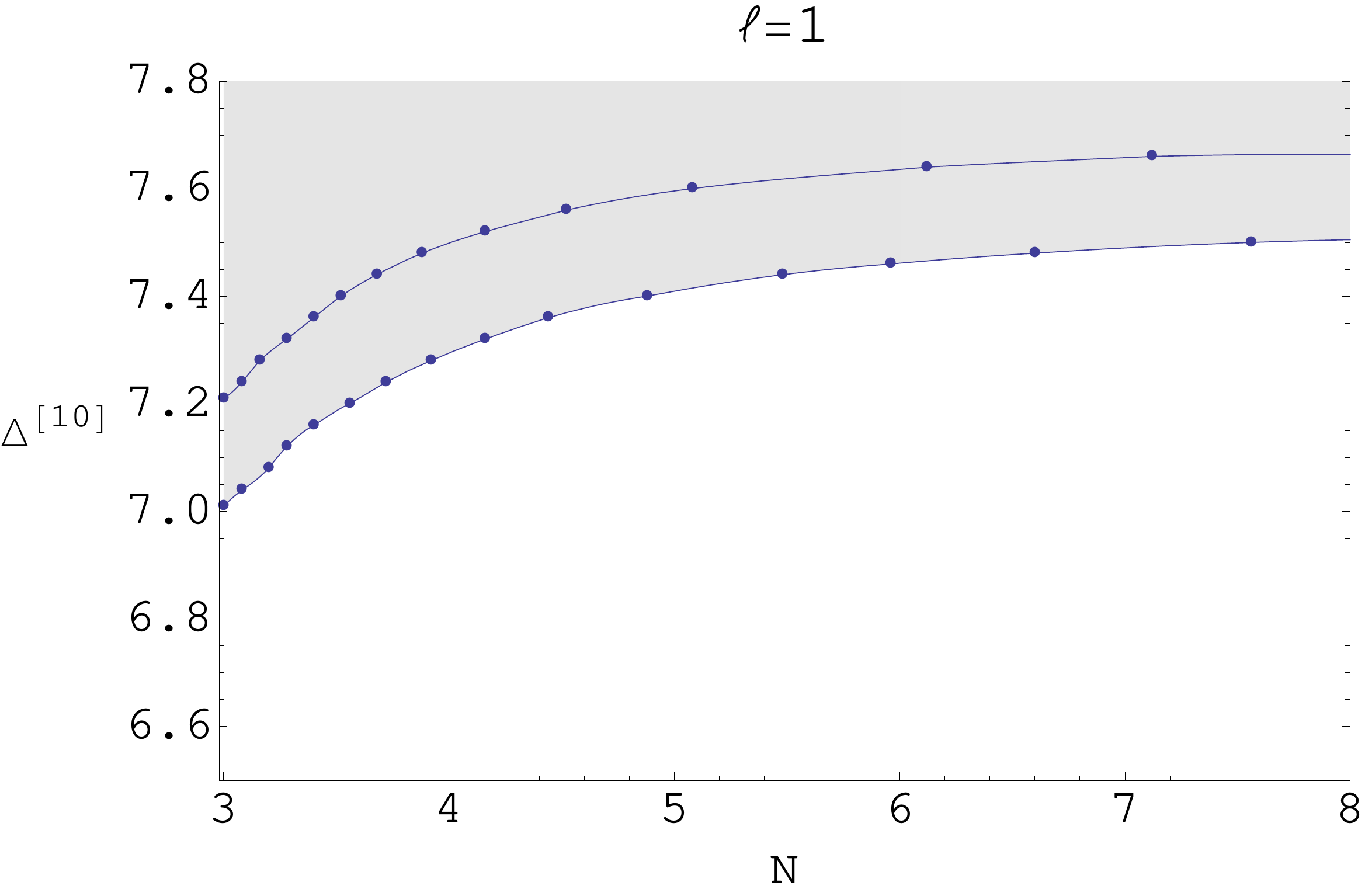}
\caption{Bounds for the scaling dimension of the leading twist unprotected superconformal primary in the $[1,0,1]$ representation of the R-symmetry group, with $\ell=1$. The given bounds correspond to $\Lambda=11$, while the upper curve shows the results for $\Lambda=9$. \label{bounds10}}
\end{figure}

\noindent The leading twist unprotected operators transforming in the $[1,0,1]$ representation are of the schematic form $\textrm{Tr} \Phi^I D^\ell \Phi^J \Phi^K \Phi^L +...$,~ $\ell=1,3,...$, where the indices $I,J,K,L$ are such  that the operator transforms in the $[1,0,1]$ representation \footnote{While there is a unique leading twist primary operator transforming in the singlet representation, this is not true for the representations $[1,0,1]$  (except for $\ell=1$) and $[0,2,0]$. In order to compute the anomalous dimensions in perturbation theory one would have to solve a mixing problem, which will include not only single trace operators, define the operators properly, etc. The leading twist operator is by definition the one with the smallest anomalous dimension. See {\it e.g.} \cite{Bianchi:2002rw} where this problem is solved to one loop for the operators with $\ell=0$ in the $[0,2,0]$ representation. } The bounds are stronger for the case with lowest spin $\ell=1$ and are shown in figure \ref{bounds10}.

\begin{figure}[h!]
\centering
\includegraphics[width=4.5in]{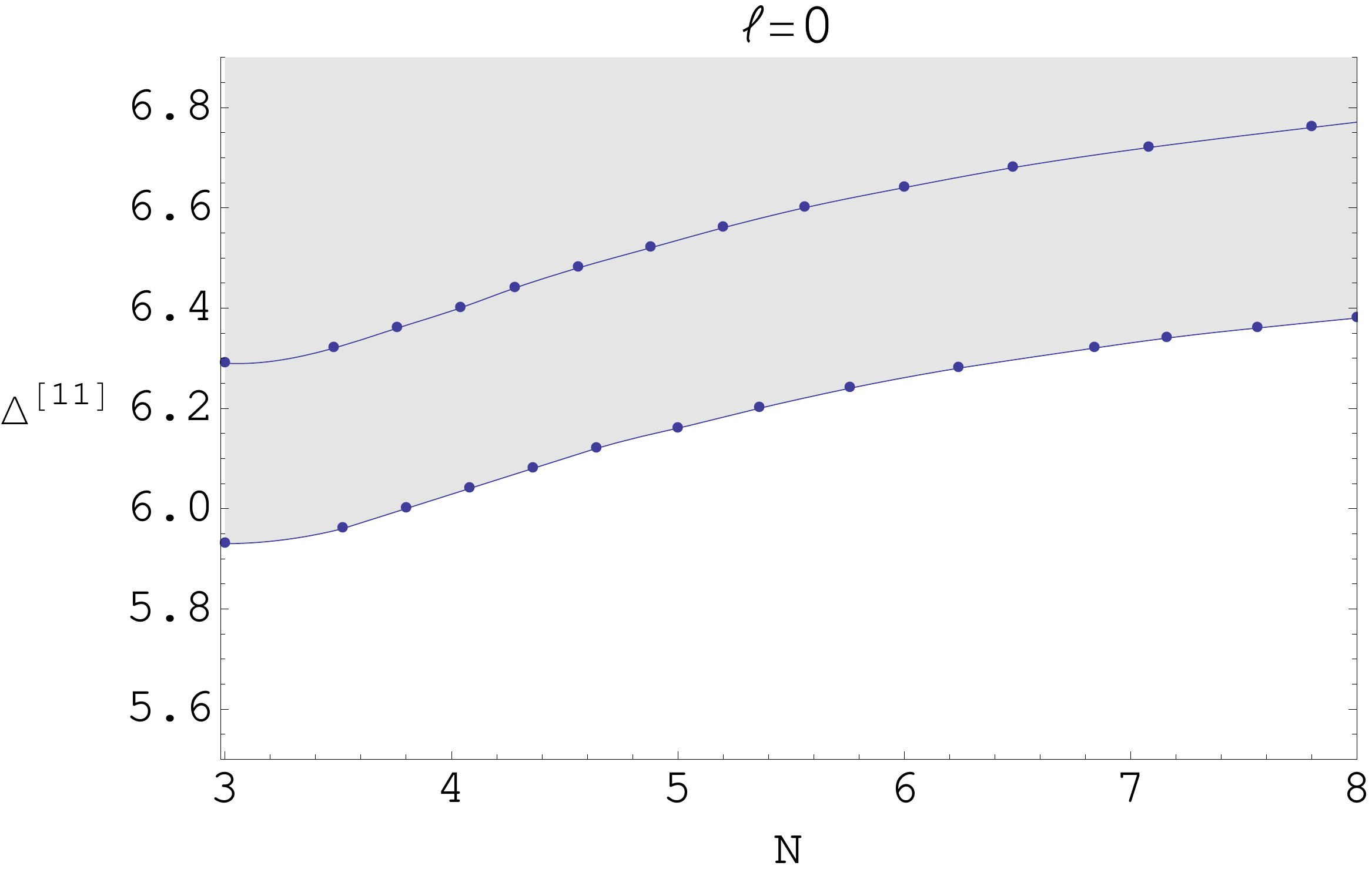}
\caption{Bounds for the scaling dimension of the leading twist unprotected superconformal primary in the $[0,2,0]$ representation of the R-symmetry group, with $\ell=0$. The stronger bounds are for $\Lambda=11$ while the upper curve corresponds to $\Lambda=9$. \label{bounds11}}
\end{figure}

\noindent The leading twist unprotected operators transforming in the $[0,2,0]$ representation are of the schematic form $\textrm{Tr} \Phi^I D^\ell \Phi^I \Phi^{(J} \Phi^{K)}+...$,~ $\ell=0,2,...$. Again, the strongest bounds are found for the case with lowest spin $\ell=0$ and are shown in figure \ref{bounds11}.

The bounds presented in this paper were obtained by using linear operators with up to 11 derivatives, or $\Lambda=11$. From the plots, we see a significative difference between these bounds and the bounds obtained with $\Lambda=9$. It seems we haven't yet exploited the full power of the bootstrap equations and one should be able to improve these bounds by increasing the number of derivatives or by more efficient methods. 

As for the singlet case, at large $N$ we expect the leading twist operators to be given by double trace operators and the dimension to behave as $\Delta \approx \Delta_0+2 - \kappa/N^2$  \cite{D'Hoker:1999jp,Arutyunov:2000ku}. Bounds in the large $N$ limit can be obtained by analyzing the bootstrap equations for $a_1=1,~a_2=a_3=0$ and we obtain $\Delta^{[10]} \lesssim 7.54$ and  $\Delta^{[11]} \lesssim 6.58$. While these bounds are a little too high at large $N$ (but, as explained above, it is expected that these bounds can be improve as we increase the number of derivatives), figures \ref{bounds10} and \ref{bounds11} show the correct behavior as we decrease $N$. It would be very interesting to compute $\kappa$ for each case by  holographic methods and compare it to our results. 

\section{Discussion}

In this paper we studied the consistency of four-point functions of half-BPS chiral primary operators of weight $p$ in four-dimensional ${\cal N}=4$ SCFT. Superconformal symmetry together with the structure of the OPE and crossing symmetry imply a set of coupled bootstrap equations. These bootstrap equations put upper bounds to the scaling dimension of unprotected superconformal primary operators transforming non-trivially under the $SU(4)$ R-symmetry group. These bounds depend not only on the central charge but also on additional parameters that appear in the OPE of two symmetric traceless tensor fields. We have analyzed in detail the case $p=3$ and found bounds for operators in the $[1,0,1]$ and $[0,2,0]$ representations for ${\cal N}=4$ SYM with gauge group $SU(N)$. These bounds represent rigorous, non-perturbative, information about non-planar ${\cal N}=4$ SYM.  

There are several possible directions on could follow. From the comparison between the bounds for $\Lambda=9$ and $\Lambda=11$ it seems one should be able to improve the bounds found in this paper, by increasing the number of derivatives or by applying more efficient methods, {\it e.g.} as in \cite{Poland:2011ey} or \cite{El-Showk:2014dwa}.

It should be straightforward to write down the bootstrap equations for the $p=4$ case. This will allow to find bounds for the dimension of operators in other representations. On the other hand, for $p>4$ one would have to work out the explicit substractions to make the conformal partial wave expansion consistent with unitarity.

It should be straightforward to extend the present bounds to bounds for the structure constants. This was done for operators in the singlet representation in \cite{Alday:2013opa}.

For the case of ${\cal N}=4$ SYM, it would be interesting to understand how S-duality acts on the above quantities. For leading twist superconformal primary operators singlets under $SU(4)$ one expects the scaling dimensions to be modular invariant \cite{Beem:2013hha}. The situation is less clear for operators transforming in non-trivial representations, such as the ones studied in this paper, since in perturbation theory one has a mixing problem. Understanding how S-duality  acts on these operators may allow, for instance, to study them in the whole fundamental region, along the lines of \cite{Beem:2013hha,Alday:2013bha}. It was conjectured in \cite{Beem:2013qxa} that the bounds from the conformal bootstrap are saturated at special values of the coupling constant. It would be very interesting to test such conjecture with non-singlets operators. 

Finally, it would be interesting to extract analytic information from the bootstrap equations of section \ref{booteq}, along the lines of \cite{Fitzpatrick:2012yx,Komargodski:2012ek,Alday:2013cwa}. For instance, following \cite{Alday:2013cwa} one may be able to understand the large spin behavior of the structure constants involving operators in non-trivial $SU(4)$ representations.

\subsection*{Acknowledgments}
\noindent
We would like to thank S. Rychkov for enlightening discussions and L. Bissi and S. El-Showk  for assistance with CPLEX. The work of the authors is supported by ERC STG grant 306260. L.F.A. is a Wolfson Royal Society Merit Award holder.

\appendix

\section{Expressions for $F_3^{[nm]}(z,\bar{z})$}
\label{fshorts}

For the case $p=3$ the suitable substractions to render the conformal wave expansion consistent with unitarity have been worked out in \cite{Dolan:2004iy}. They result in the following expressions for $F_3^{[nm]}(z,\bar{z})$
\begin{eqnarray}
F_3^{[00]}(z,\bar{z})&=&-F^{[00]}_a(z,\bar{z})+F^{[00]}_b(z,\bar{z})\\
F_3^{[11]}(z,\bar{z})&=&F^{[11]}_a(z,\bar{z})-F^{[11]}_b(z,\bar{z})\\
F_3^{[10]}(z,\bar{z})&=&F^{[10]}_a(z,\bar{z})-F^{[10]}_b(z,\bar{z})
\end{eqnarray}
where
\begin{eqnarray*}
F^{[00]}_a(z,\bar{z})&=&\sum_{\ell=0,2,\dots}\frac{2^{\ell-1} ((\ell+2)!)^2 (\ell (\ell+5) (a_1 (\ell+1) (\ell+4)-8 a_2)-12 a_2+6 a_3)}{3 (2 \ell+4)!} u G_{\ell+6}^{(\ell)}(z,\bar{z})\\
F^{[00]}_b(z,\bar{z})&=&\sum_{\ell=0,2,\dots} \left( \frac{2^{\ell-2} ((\ell+1)!)^2 (-6 a_1 (\ell-1) (\ell+1) (\ell+2) (\ell+4)+24 (2a_2+a_3)}{3 (2 \ell+2)!}\right. \\
&&\left. +\frac{24 a_2 (\ell+1) (\ell+2))}{3 (2 \ell+2)!}\right)G_{\ell+4}^{(\ell)}(z,\bar{z})\\
F^{[11]}_a(z,\bar{z})&=&\sum_{\ell=0,2,\dots}\frac{2^{\ell-2} \left(\ell^2+5 \ell+6\right) ((\ell+2)!)^2 \left(a_1 \left(\ell^2+5 \ell+4\right)+4 a_2 \right)}{3 (2 \ell+4)!} u G_{\ell+6}^{(\ell)}(z,\bar{z})\\
F^{[11]}_b(z,\bar{z})&=&\sum_{\ell=0,2,\dots}\left(\frac{2^{\ell-3} ((\ell+3)!)^2 (-6 a_1(\ell+1) (\ell+3) (\ell+4) (\ell+6)+24 (2 a_2+a_3)}{9 (2 \ell+6)!} \right.\\
&&\left.+\frac{24 a_2(\ell+3) (\ell+4))}{9 (2 \ell+6)!} u^2\right) G_{\ell+8}^{(\ell)}(z,\bar{z})\\
F^{[10]}_a(z,\bar{z})&=&\sum_{\ell=1,3,\dots}\frac{2^{\ell-1} ((\ell+3)!)^2 ((\ell+1) (\ell+6) (a_1(\ell+2) (\ell+5)-8 a_2)-12a_3+6 a_3)}{9 (2 \ell+6)!}u^2 G_{\ell+8}^{(\ell)}(z,\bar{z})\\
F^{[10]}_b(z,\bar{z})&=&\sum_{\ell=1,3,\dots} \frac{2^{\ell-1} \left(\ell^2+3 \ell+2\right) ((\ell+1)!)^2 (a_1\ell (\ell+3)+4 a_2)}{3 (2 \ell+2)!} G_{\ell+4}^{(\ell)}(z,\bar{z})
\end{eqnarray*}
In order to perform the sums note that $F^{[nm]}_{i}(z,\bar{z})$ with $i=a,b$ can be decomposed as 
\begin{equation}
F^{[nm]}_i(z,\bar{z})=f^{[nm]}_i(z,\bar{z})+f^{[nm]}_i(\bar{z},z)
\end{equation}
By using the following integral representation of the hypergeometric function
\begin{equation}
 _2F_1(a,b,c,z)=\frac{ \Gamma(c)}{\Gamma(b)\Gamma(c-b)} \int_{0}^1  \frac{t^{b-1}(1-t)^{c-b-a}}{(1-t z)^a}~dt~,
\end{equation}
It is possible to perform the sum over $\ell$ and then integrate over $t$. The final answer is
\begin{multline}
f^{[00]}_a(z,\bar{z})=-\frac{((z-2) \log (1-z)-2 z) \left((\bar{z}-2) \bar{z} \left(a_1
   \left(2 \bar{z}^4+\bar{z}^3+5 \bar{z}^2-12
  \bar{z}+6\right)\right. \right.}{z^2
   (\bar{z}-1)^3 \bar{z}(z-\bar{z})}\\
   +\frac{\left. \left.(\bar{z}-1) \left(2 a_2\left(2 \bar{z}^2+9
   \bar{z}-9\right)+3 a_3(\bar{z}-1)\right)\right)+6 (\bar{z}-1)^3
   \log (1-\bar{z}) (2 a_1+6 a_2+a_3)\right)}{z^2
   (\bar{z}-1)^3 \bar{z} (z-\bar{z})}   \end{multline}
\begin{equation}
f^{[00]}_b(z,\bar{z})=\frac{\bar{z} \log (1-z) \left(a_1\left(\bar{z}^4-2 \bar{z}^3+4
   \bar{z}-2\right)+(\bar{z}-1) \left(a_2 (\bar{z}-2)^2-a_3
   \bar{z}+a_3\right)\right)}{z ( \bar{z}-1)^3 (z- \bar{z})}
\end{equation}
\begin{equation}
f^{[11]}_a(z,\bar{z})=\frac{(z-2) z^2 ((\bar{z}-2) \log (1-\bar{z})-2 \bar{z}) \left(a_1
   \left(z^2-z+1\right)+a_2 (-z)+a_2\right)}{(z-1)^3 \bar{z}^2
   (z-\bar{z})}
\end{equation}
\begin{equation}
\begin{gathered}
f^{[11]}_b(z,\bar{z})=\frac{5 (((z-6) z+6) \log (1-z)-3 (z-2) z) \left(\bar{z} \left(a_1
   \left(\bar{z} \left(\bar{z} \left(\bar{z}^4-2 \bar{z}^3+28
   \bar{z}-74\right)+72\right)-24\right) \right. \right.}{z^3 (\bar{z}-1)^3 \bar{z}^2 (z-\bar{z})}\\
   +\frac{\left.\left.(\bar{z}-1) (a_2
   ((\bar{z}-2) \bar{z} ((\bar{z}-2) \bar{z}+48)+48)-a_3
   (\bar{z}-1) ((\bar{z}-12) \bar{z}+12))\right) \right.}{z^3 (\bar{z}-1)^3 \bar{z}^2 (z-\bar{z})}\\
  \frac{\left. -6 (\bar{z}-2)
   (\bar{z}-1)^3 \log (1-\bar{z}) (2 a_1+4
   a_2+a_3)\right)}{z^3 (\bar{z}-1)^3 \bar{z}^2 (z-\bar{z})}
   \end{gathered}
 \end{equation}
\begin{equation}
 \begin{gathered}
f^{[10]}_a(z,\bar{z})=   
  -\frac{5 \left(\left(z^2-6 z+6\right) \log (1-z)-3 (z-2) z\right)}{3 z^3 (\bar{z}-1)^3
   \bar{z}^3 (z-\bar{z})} \times \\
 \left(12
   (\bar{z}-1)^3 \log (1-\bar{z}) \left(a_1 \bar{z}^2-2 a_2
   \left(\bar{z}^2-15 \bar{z}+15\right)+3 a_3 \left(\bar{z}^2-5
   \bar{z}+5\right)\right) \right.\\
  \left. +(\bar{z}-2) \bar{z} \left(a_1 \bar{z}^2
   \left(2 \bar{z}^4+\bar{z}^3+5 \bar{z}^2-12 \bar{z}+6\right)+2
   a_2 \left(2 \bar{z}^5+7 \bar{z}^4+72 \bar{z}^3-261
   \bar{z}^2+270 \bar{z}-90\right) \right. \right.\\
 \left.  \left. +3 a_3 \left(\bar{z}^2-30
   \bar{z}+30\right) (\bar{z}-1)^2\right)\right)      
    \end{gathered}
    \end{equation}
   \begin{equation}
   f^{[10]}_b(z,\bar{z})= \frac{(\bar{z}-2) \bar{z}^2 \log (1-z) \left(a_1
   \left(\bar{z}^2-\bar{z}+1\right)+a_2
   (-\bar{z})+a_2\right)}{3 z (\bar{z}-1)^3 (z-\bar{z})}
   \end{equation}

\section{Color factors}
\label{colorfactors}

In this appendix we compute the color factors corresponding to different topologies for the case $p=3$:

\begin{figure}[h!]
\centering
\includegraphics[width=5in]{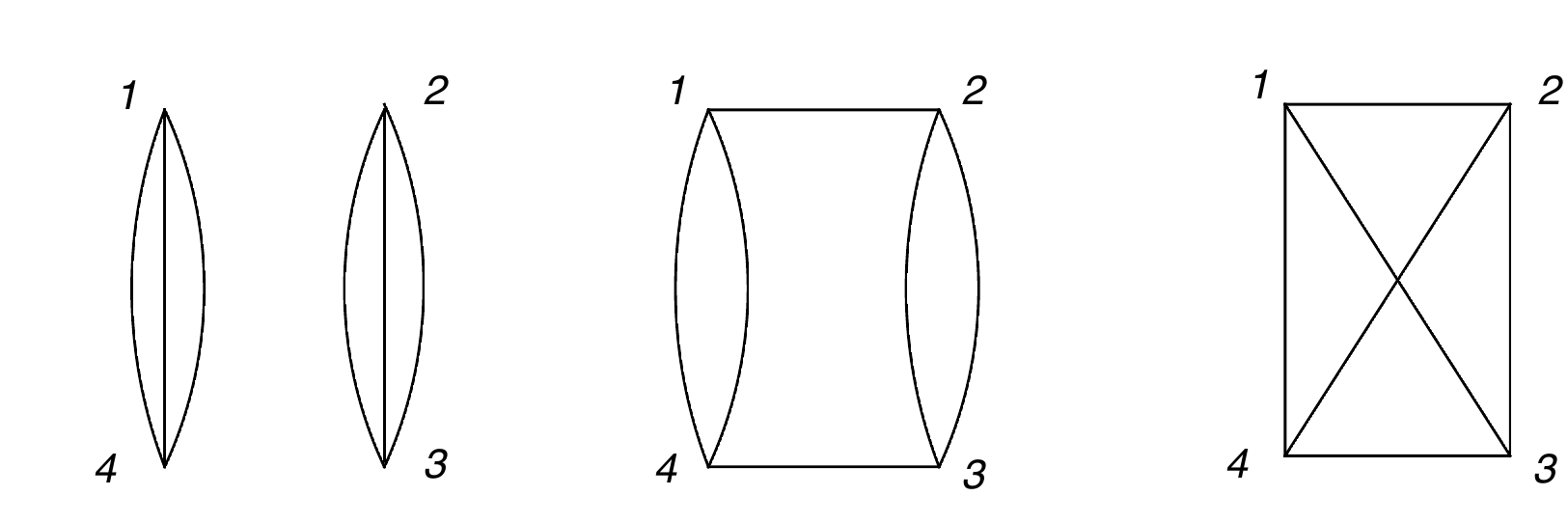}
\caption{Topologies corresponding to color factors $a_1$ (left), $a_2$ (middle) and $a_3$ (right).}
\end{figure}

We take the gauge group to be $SU(N)$. The generators $T^a$, $a=1,...,N^2-1$ and the trace satisfy  
\begin{eqnarray}
Tr(T^a T^b)=\delta^{a b},~~~Tr(1) = N
\end{eqnarray}
Contractions can be conveniently made by using the following rules
\begin{eqnarray}
\textrm{Tr}(T^a A)Tr(T^a B)= \textrm{Tr}(A B)- \frac{1}{N} \textrm{Tr} A ~Tr B,\\
\textrm{Tr} (T^a A~T^a B)= \textrm{Tr} A ~ Tr B-\frac{1}{N} \textrm{Tr}(A~B),
\end{eqnarray}
which are a consequence of $T^a_{pq}T^a_{rs}= \delta_{ps}\delta_{qr} - \frac{1}{N} \delta_{pq} \delta_{rs}$.
Let us start by computing the disconnected contribution $a_1$. There are six ways to contract $Tr \phi^3$ with itself. These split into $3+3$:

\begin{equation}
a_1= 9 (\langle abc|abc \rangle +\langle abc|bac \rangle)^2
\end{equation}
where we have introduced the notation $\langle abc|abc \rangle = \textrm{Tr}~T^a T^b T^c ~\textrm{Tr}~T^a T^b T^c$, etc. After a short calculation we obtain

\begin{equation}
\langle abc|abc \rangle=-2\frac{N^2-1}{N},~~~~~~\langle abc|bac \rangle = (N^2-1)(N-\frac{2}{N})
\end{equation}
which results in
\begin{equation}
a_1 = 9 (N^2-1)^2(N-\frac{4}{N})^2
\end{equation}
In order to compute $a_2$ it is convenient to compute the following building blocks

\begin{eqnarray}
\langle abc|abd \rangle= -\frac{2}{N}\delta^{cd},~~~~~\langle abc|bad \rangle= \left(N-\frac{2}{N} \right)\delta^{cd}
\end{eqnarray}
The symmetry factor can be counted as follows. There are 18 ways to contract two scalars in $Tr\,\Phi^3(x_1)$ to two scalars in $Tr\,\Phi^3(x_4)$. These split into $9+9$. The same is true for the other two operators. Hence

\begin{equation}
a_2= 81 \left( \left(N-\frac{4}{N} \right)\delta^{cd} \right) \left(  \left(N-\frac{4}{N} \right)\delta^{cd}  \right) =81  \left(N-\frac{4}{N} \right)^2 (N^2-1)
\end{equation}
Now we compute $a_3$. There are 162 ways to contract the scalars in $Tr\,\Phi^3(x_1)$ to one of each of the scalars of the remaining operators. For each of this possibility we have 8 contributions. Using cyclic symmetry these split into $3+4+1$:

\begin{eqnarray}
\langle abc|ade|bdf|cef \rangle &=& (N^2-1)(\frac{6}{N^2}-1)\\
\langle abc|ade|bdf|cfe \rangle &=& (N^2-1)(\frac{6}{N^2}-2)\\
\langle abc|ade|bef|cfd \rangle &=& (N^2-1)(N^2-5+\frac{6}{N^2})
\end{eqnarray}
Putting all the contributions together we obtain

\begin{equation}
a_3 = 162(N^2-1)\frac{48-16 N^2+N^4}{N^2}
\end{equation}
Let us end this appendix by computing the analog of the color factor $a_2$ for general $p$, which we denote $a^{(p)}_2$, for a generic gauge group $G$. The disconnected contribution is given by

\begin{equation}
\label{a1gen}
a^{(p)}_1= \left( \sum_{\sigma(p)} \langle a_1 a_2 ... a_p| \sigma(a_1) \sigma(a_2)...\sigma(a_p)\rangle \right)^2
\end{equation}
where the sum runs over all permutations of $p$ elements. Next, let us consider the following building block
\begin{equation}
\label{a2gen}
\kappa_{bc} \equiv \sum_{\sigma(p-1)} \langle a_1 a_2 ... a_{p-1} b | \sigma(a_1) \sigma(a_2)...\sigma(a_{p-1}) c \rangle 
\end{equation}
Due to the index structure we must have $\kappa_{bc} = \kappa \delta_{bc}$. In terms of this building block $a^{(p)}_2$ is simply given by

\begin{equation}
a^{(p)}_2 = p^4 \kappa_{bc} \kappa_{bc} = p^4 \kappa^2 \delta^{bc}\delta^{bc} = p^4 \kappa^2 ~\textrm{dim}~G
\end{equation}
In order to compute $\kappa$ consider
\begin{equation}
\kappa ~\textrm{dim}~G = \delta^{bc}\kappa_{bc}=\sum_{\sigma(p-1)} \langle a_1 a_2 ... a_{p-1} b | \sigma(a_1) \sigma(a_2)...\sigma(a_{p-1}) b \rangle = \frac{1}{p} \sqrt{a_1}
\end{equation}
We arrive at the final expression
\begin{equation}
a^{(p)}_2= \frac{p^2}{\textrm{dim}~G }~a^{(p)}_1
\end{equation}

\end{document}